\documentclass[prl, twocolumn, showpacs]{revtex4}
\usepackage{amssymb}
\usepackage{amsmath}
\usepackage{epsfig}
\usepackage{bm}

\begin{document}

\title{Dynamic spin-charge coupling: \\
\text{\boldmath$ac$} spin Hall magnetoresistance in non-magnetic conductors}

\author{ P.S. Alekseev$^1$ and M.I. Dyakonov$^2$ }

\affiliation{$^{1}$Ioffe Institute,194021, St. Petersburg, Russia\\
$^{2}$Laboratoire Charles Coulomb, Universit\'e Montpellier, CNRS, France}

\begin{abstract}

The dynamic coupling between spin and charge currents in non-magnetic conductors is considered.
As a consequence of this coupling, the spin dynamics is directly reflected in the electrical impedance
of the sample, with a relevant frequency scale defined by spin relaxation and spin diffusion. This allows
the observation of the electron spin resonance by purely electrical measurements.

\end{abstract}

\maketitle

{\em 1. Introduction.}
It was predicted nearly half a century ago \cite{DP1, DP2} that spin-orbit
interaction results in the interconnection between electrical and spin currents:
an electrical current produces a transverse spin current and {\it vice versa}.
This leads respectively to the direct and inverse spin Hall effects. Following the
proposal in Ref.~\cite{Averkiev}, the inverse spin Hall effect was observed
experimentally by Bakun {\it et al.}~\cite{Bakun} in 1984, without causing much
excitement at that time.

Twenty years later, after the first experimental observations of the (direct) spin Hall
effect \cite{Kato, Wunderlich} this topic has become a subject of considerable interest
with thousands of publications, see for example a review in Ref.~\cite{DKh}.

Because of the interconnection between the spin and charge currents, anything that
happens with spins will influence the charge current, i.e. result in corresponding changes
of the electrical resistance, which can be measured with a very high precision. An example
of this link is provided by the {\it spin Hall magnetoresistance} \cite{Dyakonov}, the
reason for which is the depolarization of spins accumulated at the sample boundaries by
a transverse magnetic field and the resulting decrease of the driving electric current
(for a given voltage) \cite{rem}. This effect was experimentally demonstrated in platinum
by V\'elez {\em et al.} \cite{Velez}.

Earlier, a similar effect was discovered and studied by Nakayama {\it et al.} \cite{Nakayama} in
layered structures ferromagnet-normal metal. The magnetization in the ferromagnet can be
rotated by an applied magnetic field which results in a change in the normal metal resistivity.

In recent years, the {\it ac} spin Hall effect in ferromagnet-normal metal structures has also been
studied both experimentally \cite{Wei, Hyde, Weiler, Hahn} and theoretically \cite{Chiba, Chen, Ulloa}.
The precession of the magnetization in a ferromagnet leads to a time-varying injection of spin into
the normal metal. Due to the inverse spin Hall effect, the resulting spin current in the normal metal
generates the {\it ac} electric current.

In particular, the observed {\it ac} voltage  resonantly depends on the Larmor frequency in the
ferromagnet and the frequency of the external {\it ac} magnetic field, which excites the precession
of magnetization. In this way, with the aid of the spin Hall effect in a normal metal, the ferromagnetic
resonance was observed by electric measurements.

While these studies are quite important for achieving the ultimate goal of storing and manipulating
information by the use of spin Hall effect for switching magnetic domains in magnetics (see the
reviews \cite{Hoffmann1, Hoffmann2}), the physics of the layered magnetic structures is quite
complicated, and this makes the exact theory and the quantitative analysis of experimental data
rather difficult.

\begin{figure}[b]
\centerline{\includegraphics[width=1\linewidth]{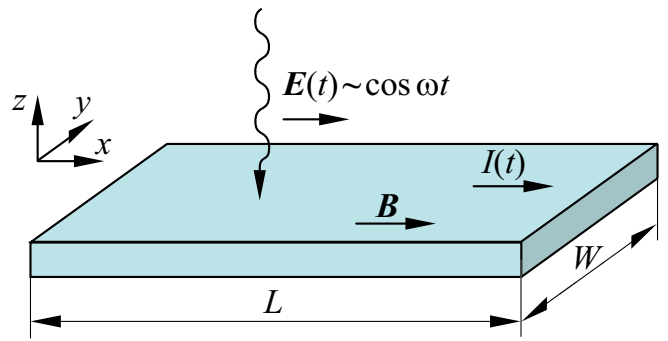}}
\caption{ A metal or semiconductor sample in a magnetic field $\boldsymbol{B}$ and {\em ac} electric
field  $\boldsymbol{E}(t)$. The length of the sample, $L$, is much greater than its width, $W$.
It is assumed that the {\em ac} electric field penetrates everywhere into the sample, i.e. that the
electron system is either two-dimensional, or three-dimensional, but with thickness less than the skin depth.}
\end{figure}

Here, following Ref.~\cite{Dyakonov},  we develop a much more simple theory of {\it ac} electron
magnetotransport controlled by the direct and inverse spin Hall effects in {\it non-magnetic}
materials, semiconductors or metals. The theory is based on the phenomenological transport equations
\cite{DP1, DP2, Dyakonov, DKh} describing the interconnection between spin and charge currents. We show
that spin resonance in non-magnetic materials can be observed by purely transport measurements.

{\em 2. Transport equations.}
Consider a conductor in an external {\it ac} electric field $\boldsymbol{E}(t)\sim \cos(\omega t)$ and
a magnetic field $\boldsymbol{B}$, see Fig. 1.  We assume that the {\it ac} frequency $\omega$ is much
lower than the cyclotron frequency $\omega_c$, and that $\omega\tau \ll 1$, where $\tau$ is the momentum
relaxation time. However the spin Larmor frequency $\varOmega$ and the spin relaxation time $\tau_s \gg \tau$
are such that $\omega \sim \varOmega \sim 1/\tau_s$.

In this frequency range, the basic phenomenological equations for the electron flow density
$\boldsymbol{q}=\boldsymbol{j}/e$, the spin current density tensor $q_{ij}$, and the spin density
vector $\boldsymbol{P}$ are \cite{rem1,DKh}:
\begin{equation}
\label{main_eqs}
\boldsymbol{q} = \mu n \boldsymbol{E} + \gamma D \: \mathrm{rot} \boldsymbol{P}
 \: ,  \qquad \qquad  \quad \;
\end{equation}
\begin{equation}
\label{main_eqs_qij}
 q_{ij} = -D \frac{\partial P_j }{\partial x_i} + \gamma  \mu n \: \epsilon_{ijk} E_k
 \: , \quad  \quad \;
\end{equation}
\begin{equation}
\label{main_eqs_cont}
\frac{\partial P_j }{\partial t} + \frac{\partial q_{ij} }{\partial x_i} +
 ( \boldsymbol{ \varOmega} \times \boldsymbol{  P }) _j  + \frac{P_j}{\tau_s} = 0
 \: ,
\end{equation}
 where $n$ is the electron density, $\mu$ is the electron mobility, $D$  is the diffusion
coefficient, $\gamma \ll 1$ is the dimensionless parameter proportional to the strength
 of the spin-orbit interaction and describing the interconnection between the particle
and the spin currents, $\epsilon_{ijk}$ is the unit antisymmetric tensor, the vector
 $\boldsymbol{\varOmega}$ is directed along the applied magnetic field, $\varOmega$ being
the Larmor frequency for electron spins, and $\tau_s$ is the spin relaxation time.

The first term  in Eq.~(\ref{main_eqs}) is the usual Drude contribution,
 while the second term expresses the interconnection (caused by spin-orbit interaction)
between particle current and  the spin current caused by the inhomogeneity of the spin density.

Eq.~(\ref{main_eqs_qij})  describes two contributions to the spin  current density $q_{ij}$:
the first one is due to diffusion of  spin-polarized electrons and the second one is  due
to the transformation of particle current into spin current.

Eq.~(\ref{main_eqs_cont}) is the continuity equation for the spin density, taking into
account spin diffusion, rotation of spin in magnetic field, and spin relaxation.

In the geometry of Fig. 1, both the particle and the spin flows depend on the $y$ coordinate only.
We consider that the {\em ac} electric field, as well as the magnetic field, are directed along the
{\em x} direction.

As it follows from Eq.~(\ref{main_eqs_qij}), the two nonzero components of the spin current
density tensor are:
\begin{equation}
 \label{q_comps__case_1}
q_{yy} = - D \frac{dP_y}{dy}
\:, \quad \;\;
\displaystyle
q_{yz} = - D \frac{dP_z}{dy} + \gamma \mu n E \cos\omega t
 \:.
\end{equation}

The absence of spin currents across the sample boudaries is described by the
boundary conditions: $q_{yy} =0$ and $ q_{yz}=0 $ at $y = \pm  W/2$.

{\em 3. Solution of the transport equations.}
 The nonzero components of the spin density vector $\boldsymbol{P}$  can be written in a complex form:
$P_y(y,t) = [P_y(y)e^{-i \omega t } + c.c.]/2$, and similarly for $P_z(y,t)$.
Then, from Eqs.~(\ref{main_eqs_cont}) and (\ref{q_comps__case_1}) we obtain a system of coupled
equations for $P_y(y)$ and $P_z(y)$:
\begin{equation}
 \label{exact_eqs}
\begin{array}{l}
\displaystyle
D \frac{d^2 P_y}{dy^2} = \Big(-i \omega + \frac{1}{\tau_s} \Big) P_y +\varOmega P_z
\:,
\\
\\
\displaystyle
D \frac{d^2 P_z}{dy^2} = \Big(-i \omega + \frac{1}{\tau_s} \Big) P_z -\varOmega P_y
\:,
\end{array}
\end{equation}
with the boundary conditions:
\begin{equation}
 \label{bound_conds}
\left. \frac{d P_y}{dy} \right| _ {y=\pm W/2} = 0
\:, \quad  \;\; \;
\left. \frac{d P_z}{dy} \right| _ {y=\pm W/2} = \gamma \frac{\mu nE}{D}
\:.
\end{equation}
The solution of Eqs.~(\ref{exact_eqs}) with the boundary conditions (\ref{bound_conds})
 yields the spin density profile:
\begin{equation}
\label{P(y)}
\begin{array}{c}
 \displaystyle
P_y(y)
 = -i\gamma
 \frac{\mu n E}{2D} \,
 [ \, F_+ (y) - F_- (y) \, ],
\\
\\
 \displaystyle
P_z(y)
 = \gamma \frac{\mu n E}{2D} \, [ \,  F_+ (y) + F_- (y)\, ],
 \end{array}
\end{equation}
 where
\begin{equation}
\label{F}
F_{\pm}(y) = \frac{\sinh (\lambda _{\pm} y) }{\lambda _{\pm} \cosh (\lambda _{\pm}W/2)}
 \: ,
\end{equation}
\begin{equation}
\label{lambda}
\lambda_{\pm} = \frac{ \sqrt{1+ i (-\omega \pm \varOmega )\tau_s}}{ L_s}
\:,
\end{equation}
and $L_s = \sqrt {D \tau_s}$ is the spin diffusion length.

Thus for narrow samples, $|\lambda_{\pm}|W \ll 1$, the spin density $\boldsymbol{P}$
depends linearly on the coordinate $y$, while for wide samples, $|\lambda_{\pm}|W \gg 1$,
the spin density $\boldsymbol{P}$  is concentrated near the sample edges.
In the last case, spin density exhibits spin resonance at $\omega = \pm \varOmega$
provided that $\omega \tau_s \gtrsim 1$. The signs $\pm$ correspond to the
contribution to $\boldsymbol{P}$ from the components of $\boldsymbol{E}(t)$
with the right  and the left circular polarizations, respectively.

The current density $j_x=eq_x$ can now be calculated using Eq. (\ref{main_eqs}):
$j_x  = e\mu n E \cos \omega t + \Delta j(y,t)$, where the first term is the normal
Drude contribution (in the assumed limit $\omega\tau \ll 1)$, while the second term
is a correction which is of second order in the spin-orbit interaction:
$ \Delta j(y,t) = [\Delta j(y) e^{-i \omega t } + c.c. ]/2$, where
\begin{equation}
 \label{Delta_q}
 \Delta j(y) = \gamma^2 \frac{e\mu n E}{2} \, \Big[\,\frac{ dF_{+}}{dy} +
 \frac{ dF_{-}}{dy}\, \Big]\:.
\end{equation}
For wide samples, $| \lambda _{\pm} | W \gg 1 $, this correction to the {\em ac} current
density, like the spin density, is concentrated near the sample edges.

The spin-orbit correction $\Delta I$ to the main Drude part, $I_0 = e\mu n E W $,
of the total current can be calculated from Eq.~(\ref{Delta_q}):
\begin{equation}
\label{I}
 \Delta I = \int_{-W/2} ^{W/2} \Delta j(y) \, dy
 \:.
\end{equation}
 Thus we obtain the final result for the correction to the sample impedance,
 $\Delta Z = \Delta Z (\omega,\varOmega)$, caused by spin-orbit interaction:
\begin{equation}
\label{Delta_I_exañt}
\frac{\Delta Z}{Z_0}  =   -\gamma^2    \Big[
 \frac{ \tanh ( \lambda_+ W/2) }{\lambda_+ W}
+
\frac{ \tanh ( \lambda_- W/2)}{\lambda_- W}
 \Big]
\:,
\end{equation}
where $Z_0 = L/(e\mu n W)$ and $\lambda _{\pm}$ are defined by Eq.~(\ref{lambda}).

{\em  4. Results and discussion.}
We now analyze our results given by Eq. (\ref{Delta_I_exañt}) for some special cases.

{\em (i) Low frequencies: $\omega\tau_s \ll 1 $.  }
The results coincide with those of Ref.~\cite{Dyakonov}
for the {\em dc} spin Hall magnetoresistance.

\begin{figure}[t!]
\centerline{\includegraphics[width=1\linewidth]{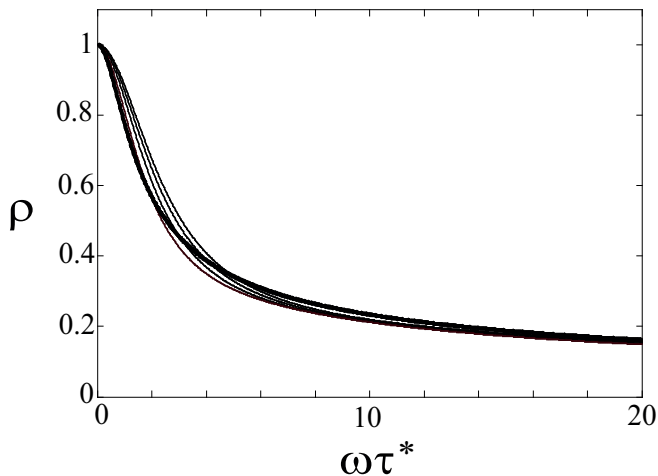}}
\caption{ The ratio $\varrho$ of the real part of the spin-orbit correction
 to impedance  $\Delta Z$  to its value at zero frequency
$\Delta Z _0   $ as a function of ac frequency $\omega$ in the absence of magnetic field.
 Thin curves  correspond to  $W/L_s = 0.2, \, 0.8,  \, 1.3,  \, 2 $, while the thick curve
corresponds to $W/L_s = \infty$. It is seen that all curves practically coincide.}
\end{figure}

{\em (ii) Zero magnetic field, $B = 0 $.}
 In Fig.~2 we plot the ratio $\varrho(\omega)=\mathrm{Re} \Delta   Z (\omega,0)/\Delta Z_0$
of the real part of the spin-orbit correction $\Delta Z$ [Eq.~(\ref{Delta_I_exañt})]
in the absence of magnetic field to its value at zero frequency,
$\Delta Z _0  = \Delta Z(0, 0) $.
It is seen that $\varrho (\omega) $ has a quasi-universal behavior as a function
of the parameter $\omega \tau^* $.  Here $ 1/\tau^* = 1/\tau_s + 1/\tau_d $ is
the effective total relaxation rate which is the sum of the bulk spin relaxation rate
$1/\tau_ s $ and the diffusion rate for space inhomogeneity in the spin
distribution $1/\tau_s =  4 D /W^2$.

Thus there are the two relaxation processes: the bulk spin relaxation
with the rate $1/ \tau_s$ and the decay of spin inhomogeniety due to diffusion of spin-polarized
electrons. The quasi-universal results in Fig.~2 are quite similar
to those obtained in Ref.~\cite{Dyakonov} for the {\em dc} spin Hall magnetoresistance.

Indeed, from Eq.~(\ref{Delta_I_exañt}) one can obtain the relation between the corrections to
the {\em ac} impedance in zero magnetic field and to the {\em dc} magnetoresistance:
\begin{equation}
 \label{relation}
\mathrm{Re} \,  \Delta Z (\omega  , 0) = \Delta Z (0 ,\varOmega = -\omega  )
 \:.
\end{equation}
Here $\Delta Z (0 ,\varOmega ) $ is, in fact, the spin Hall magnetoresistance
calculated in Ref. \cite{Dyakonov} and denoted therein as $\Delta R (\varOmega )$.

{\em (iii) High frequencies: $\omega\tau_s \gg 1 $.  }
For narrow samples ($ W \ll L_s /\sqrt{\omega \tau_s}$) the correction $\Delta Z$
depends neither on frequency, nor on magnetic field in the main order by the parameter
$W/L_s \ll 1$ (at not too high magnetic fields when $\varOmega \lesssim \omega $).
 With the small correction on the order of $(W/L_s)^2$ included, we obtain:
\begin{equation}
  \label{narrow}
\frac{ \Delta Z  }{Z_0}  = -\gamma^2 \Big( \, 1- \frac{ 1- i \omega \tau_s }
 { 24 }  \frac{ W^2 } { L_s^2 } \, \Big)
\:.
\end{equation}

\begin{figure}[t!]
\centerline{\includegraphics[width=1\linewidth]{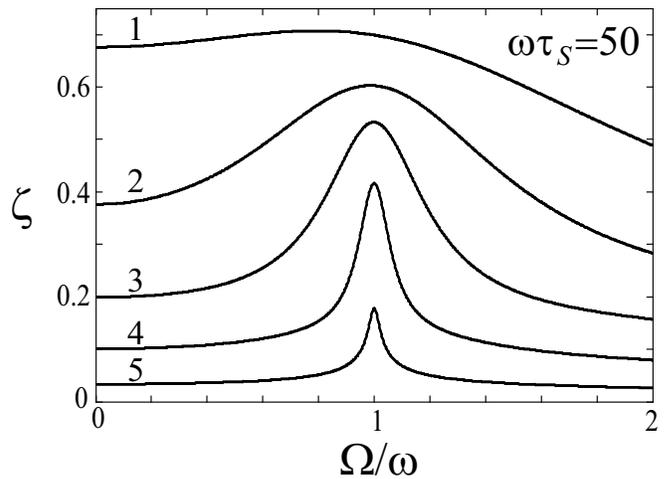}}
\caption{ The real part of the correction to the {\em ac} impedance~$\zeta $
 as a function of the Larmor frequency $ \varOmega $
 at a fixed  {\em ac} frequency $\omega$
 for medium and large sample widths.
 For curves 1, 2, 3, 4, 5 the parameter
 $W/L_s$ is equal to 0.4, 0.6,  1,  2, 6, respectively. }
\end{figure}

For wide samples ($W \gg L_s$) Eq.~(\ref{Delta_I_exañt}) leads to the formula:
\begin{equation}
  \label{wide}
\frac{\Delta Z }{Z_0} = -\gamma^2
\frac{L_s }{W}
 \sum \limits_{\pm} \frac{1}{\sqrt{1 + i (- \omega \pm \varOmega) \tau_s }}
\:,
\end{equation}
displaying spin resonance at $ \omega = \pm \varOmega$.

The general formula (\ref{Delta_I_exañt}) is needed for medium sample widths
($ L_s /\sqrt{\omega \tau_s}  \ll  W  \ll   L_s $).  In this case
Eq.~(\ref{Delta_I_exañt}) describes the crossover between the resonant dependence of
$ \Delta Z  $ on $ \varOmega $ for wide samples [Eq.~(\ref{wide})] and
the non-resonant dependence of $ \Delta Z  $ on $ \varOmega $  for narrow samples
[Eq.~(\ref{narrow})].

In Fig.~3 we plot the ratio $\zeta =\mathrm{Re} \,\Delta Z/(-\gamma^2 Z_0 )$
of the real part of the spin-orbit correction  $\Delta Z$
[Eq.~(\ref{Delta_I_exañt})] to its value for narrow samples
[Eq.~(\ref{narrow})]  as a function of $\varOmega$ at a fixed $\omega$
for different sample widths $W$.  The transition from the non-resonant
to the resonant behavior  of $\zeta (\varOmega)$ with the increase of $W$
is clearly seen. Eq.~(\ref{wide}) and Fig.~3 show that the wider is
the sample, the  smaller are both the amplitude and the width of the resonance
peak.

It is interesting to study the behavior of the normalized {\em ac} magnetoresistance
\begin{equation}
 \varrho(\omega,\varOmega)  = \frac {\mathrm{Re} \,  \Delta Z (\omega,\varOmega)}
  {\mathrm{Re} \, \Delta Z (\omega,\varOmega = \omega)}
 \:.
\end{equation}
An analysis similar to that performed above for the dependence $\varrho(\omega)$ in
zero magnetic field,  shows that $\varrho(\omega,\varOmega)$ at a fixed $\omega$ has
a quasi-universal  behavior as a function of the parameter $ (\varOmega-\omega)\tau^*$
at $ | \varOmega | > \omega $,  similar to the behavior of $\varrho (\omega)$
displayed at Fig.~2. However, the dependencies of $ \varrho $ on $\Omega $ at
$|\varOmega | < \omega $, as it is seen from Fig.~3, are qualitatively different
for wide and for narrow samples.

{\em 4. Conclusion.}
We have shown that the combination of the direct and inverse spin Hall effects
in nonmagnetic metals and semiconductors offers an interesting possibility to study high-frequency
spin phenomena, including spin resonance, by purely electrical measurements. The corresponding
corrections to the sample impedance are of second order in the spin-orbit coupling parameter,
$\gamma$.

In the absence of external magnetic field, the frequency dependence of the electrical impedance is
defined by the sum of the bulk spin relaxation rate and the spin diffusion rate. The interplay between
the two corresponding relaxation times defines also the width and the amplitude of the electrically
measured spin resonance.


\begin{thebibliography}{}

\bibitem{DP1} M.I.~Dyakonov and V.I.~Perel, JETP Lett. {\bf 13}, 467 (1971)

\bibitem{DP2} M.I.~Dyakonov and V.I.~Perel, Phys. Lett. {\bf A35}, 459 (1971)

\bibitem{Averkiev} N.S.~Averkiev and M.I.~Dyakonov, Sov. Phys. Semicond. {\bf 17}, 393 (1983)

\bibitem{Bakun} A.A.~Bakun, B.P.~Zakharchenya, A.A.~Rogachev, M.N.~Tkachuk, and V.G.~Fleisher,
JETP Lett. {\bf 40}, 1293 (1984)

\bibitem{Kato} Y.K.~Kato, R.C.~Myers, A.C.~Gossard, and D.D.~Awschalom, Science {\bf 306}, 1910 (2004)

\bibitem{Wunderlich} J.~Wunderlich, B.~Kaestner, J.~Sinova, and T.~Jungwirth, Phys. Rev. Lett. {\bf 94}, 047204 (2005)

\bibitem{DKh} M.I.~Dyakonov and A.V.~Khaetskii, In: {\it Spin Physics in Semiconductors}, M.I. Dyakonov, editor,
 2nd edition, Springer (2017), Ch. 8

\bibitem{Dyakonov} M.I.~Dyakonov, Phys. Rev. Lett. {\bf 99}, 126601 (2007)

\bibitem{rem} In Ref. \cite{Dyakonov} a 2D conductor in the {\it xy} plane was considered,
which can be also viewed as a cross-section of a 3D sample. Since there is no dependence
on the {\it z} coordinate, the results are equally applicable for a 3D conductor.

\bibitem{Velez} S. Velez, V. N. Golovach, A. Bedoya-Pinto, M. Isasa, E. Sagasta, M. Abadia,
C. Rogero, L. E. Hueso, F. S. Bergeret, and F. Casanova, Phys. Rev. Lett. {\bf 116}, 016603 (2016)

\bibitem{Nakayama} H. Nakayama, M. Althammer, Y.-T. Chen, K. Uchida, Y. Kajiwara, D. Kikuchi, T. Ohtani,
S. Geprags, M. Opel, S. Takahashi, R. Gross, G. E. W. Bauer, S. T. B. Goennenwein, and E. Saitoh,
Phys. Rev. Lett. {\bf 110}, 206601 (2013)

\bibitem{Wei} D. Wei, M. Obstbaum, M. Ribow, C. H. Back, and G. Woltersdorf,
Nat. Comm. {\bf 5}, 3768 (2014)

\bibitem{Hyde} P. Hyde, Lihui Bai, D. M. J. Kumar, B. W. Southern, C.-M. Hu, S. Y. Huang,
B. F. Miao, and C. L. Chien, Phys. Rev. B {\bf 89}, 180404 (2014)

\bibitem{Weiler} M. Weiler, J. M. Shaw, H. T. Nembach, T. J. Silva, Phys. Rev. Lett. {\bf 113}, 157204 (2014)

\bibitem{Hahn} C. Hahn, G. de Loubens, M. Viret, O. Klein, V. V. Naletov, J. Ben Youssef,
Phys. Rev. Lett. {\bf 111}, 217204 (2013)

\bibitem{Chiba} T. Chiba, G. E. W. Bauer, and S. Takahashi,
Phys. Rev. Applied {\bf 2}, 034003 (2014)

\bibitem{Chen} W. Chen, M. Sigrist, J. Sinova, D. Manske, Phys. Rev. Lett.  {\bf 115}, 217203 (2015)

\bibitem{Ulloa} C. Ulloa and R. A. Duine, Phys. Rev. Lett. {\bf 120}, 177202 (2018)

\bibitem{Hoffmann1} A. Hoffmann, IEEE Transactions on Magnetics {\bf 49}, 5172 (2013)

\bibitem{Hoffmann2} M.B. Jungfleisch, W. Zhang, R. Winkler, and A. Hoffmann,  In: {\it Spin Physics in
Semiconductors}, M.I. Dyakonov, editor, 2nd edition, Springer (2017), Ch. 11

\bibitem{rem1} In the absence of inversion symmetry there may be additional terms
describing this coupling. In particular, there is a spin current induced by a non-equilibrium
spin polarization and a uniform spin polarization generated by electric current. Here, such
effects are not considered.

\end{thebibliography}
\end{document}